\documentclass[twocolumn,showpacs,aps,floatfix]{revtex4}
\usepackage{amsmath}
\usepackage{graphicx}
\usepackage{dcolumn}
\usepackage{bm}
\begin{document}
\title{Waveplate retarders based on overhead transparencies}

\author{Igor Savukov and Dmitry Budker}
\email{budker@berkeley.edu} \affiliation{Department of Physics,
University of California, Berkeley, CA 94720-7300}
\affiliation{Nuclear Science Division, Lawrence Berkeley National
Laboratory, Berkeley CA 94720}

\begin{abstract}
We describe procedures for constructing inexpensive waveplates of
desired retardation out of ordinary commercially available
transparencies. Various relevant properties of the transparencies
are investigated: the dependence of retardation on rotation of the
film, tilt, wavelength, position, and temperature. Constructing
waveplates out of combinations of transparency sheets is also
explored.
\end{abstract}


 \maketitle
\date{\today}

\section{Introduction}
\begin{figure}
\centerline{\includegraphics*[scale=0.9]{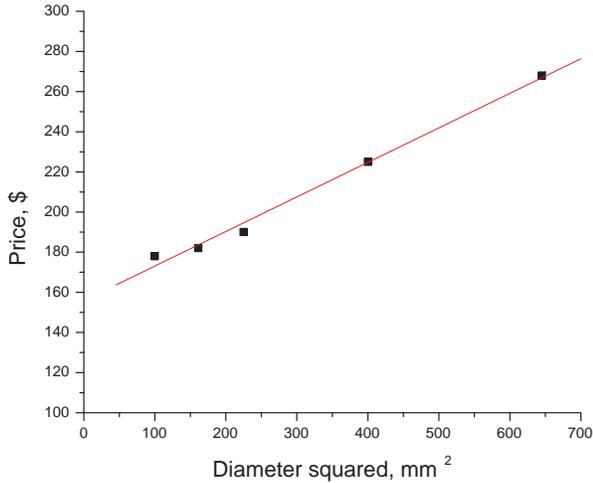}} \caption{The
price of waveplates vs size. The data are from a CASIX catalog.
Price grows linearly with the square of the size of the waveplate.}
\label{price}
\end{figure}
\begin{figure}
\centerline{\includegraphics*[scale=0.5]{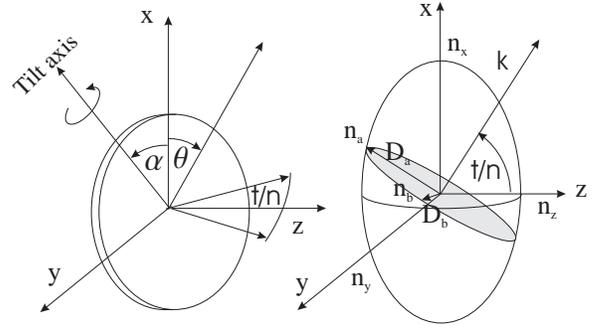}}
\caption{\textbf{On the left}: Definition of the coordinate system
of a transparency waveplate. The principal axes x and y of the index
ellipsoid, shown on the right, lie in the plane of transparency
film, while the z axis is perpendicular to this plane. The
``rotation'' is defined as the rotation around the z axis through
angle $\theta$. The tilt is defined as the rotation around the tilt
axis oriented at some angle $\alpha$ with respect to x in the
transparency plane. \textbf{On the right}: The index ellipsoid and
its intersection with the plane perpendicular to a wave normal $k$.
The intersection is an ellipse whose half-axes $n_a$ and $n_b$ are
responsible for the retardation produced by the waveplate. The tilt
around y (the case when $\alpha=90^\circ$)  is illustrated.}
\label{indell}
\end{figure}
\begin{figure}
\centerline{\includegraphics*[scale=1.0]{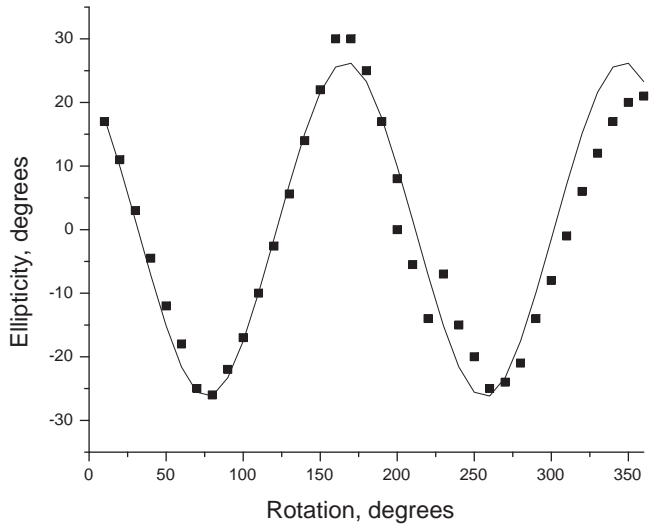}} \caption{
Ellipticity as a function of transparency rotation. The data shown
with squares  agree well with a fit using Eq. (\ref{chigen}) (the
solid line), in which $\delta=0.91$ rad and an arbitrary offset in
the rotation angle is allowed. Some aperiodicity can be observed due
to the fact that under rotation slightly different regions of the
transparency are probed. The rotation of the transparency is the
most convenient way to adjust ellipticity to a desirable level
between maximal and minimal values.} \label{transprot}
\end{figure}
\begin{figure}
\centerline{\includegraphics*[scale=1.0]{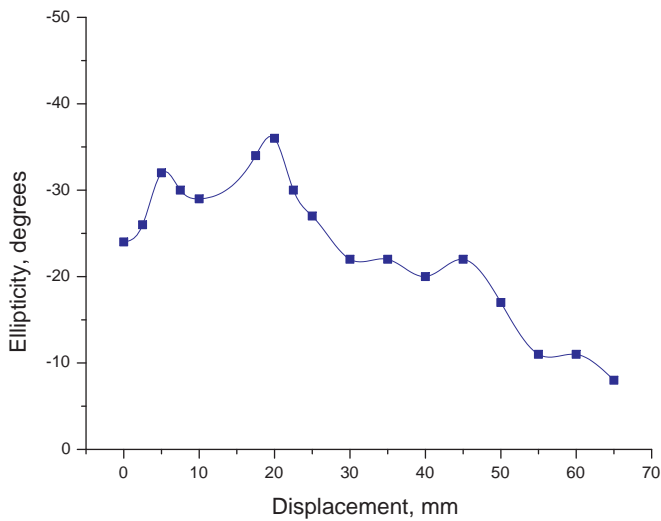}} \caption{
Ellipticity as a function of transparency translation. The size of
probing laser beam is approximately 2 mm. The data exhibit smooth
behavior with typical continuous variation 5 degrees per 10 mm and
ellipticity changing slowly in a large range. From applications
point of view, this means that waveplates of given retardation and
of the size 10-30 mm depending retardation tolerance can be made by
selecting an appropriate region. This behavior is similar for
different transparencies.} \label{trans1pos}
\end{figure}

\begin{figure}
\centerline{\includegraphics*[scale=0.26]{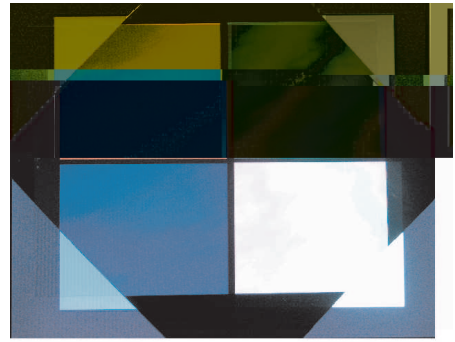}}
\caption{Large-scale retardation uniformity of a transparency sheet
in two dimensions. The transparency is placed between a 17" thin
film transistor - liquid crystal display (TFT-LCD) computer monitor,
source of polarized light with polarization at $45^\circ$ with
respect to vertical, and a square polaroid film of similar size
oriented to extinguish light in the absence of transparency.
Transparency has its edges aligned with the screen. The color
segments are generated in the computer by drawing filled rectangles
in a drawing program. Three ``pure'' color quadrants provide
information on uniformity at different wavelengths. In general it is
expected that birefringence variation caused by non-uniformity of
stretching should be similar for the three different colors,
although somewhat smaller for red.} \label{comptr1}
\end{figure}

\begin{figure}
\centerline{\includegraphics*[scale=0.6]{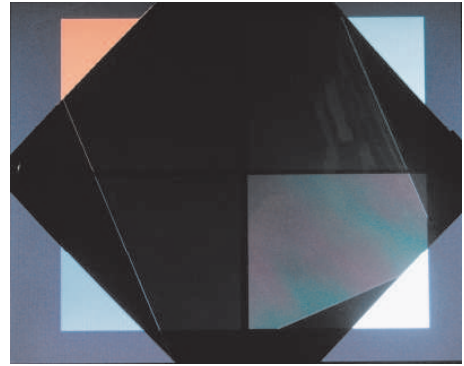}} \caption{The
uniformity of the zero-ellipticity across large area of a
transparency film, meaning that the orientation of the index
ellipsoid does not depend on position.} \label{comptr2}
\end{figure}

\begin{figure}
\centerline{\includegraphics*[scale=1.0]{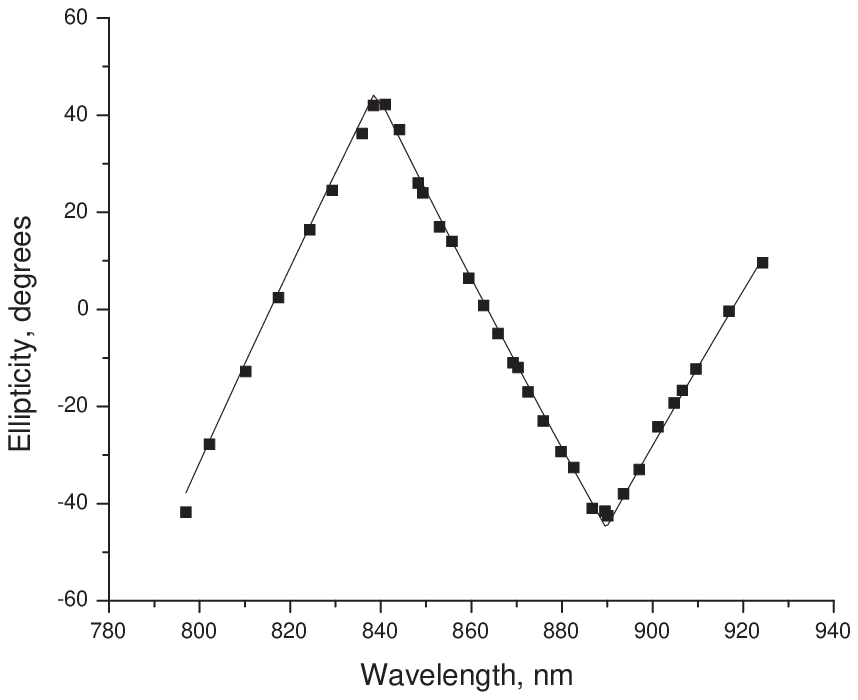}} \caption{
Ellipticity as a function of light wavelength for 3M black and white
transparency. The data follow Eq. (\ref{chigen}) with
$\theta=45^\circ$ and $\delta[rad]=46120/\lambda$[nm]. At 840 nm,
the waveplate retardation is $17.5\pi$. } \label{transwavelength}
\end{figure}
Conventionally, retarders are made of calcite, quartz, or mica
\cite{bookopt1,bookopt2, bookopt3}. In applications of wave plates
in which low price and rapid availability are more important than
high quality, the construction of wave plates from inexpensive
materials can be of great practical interest. For example, ordinary
cellophane used in commercial packaging can be very close to a
half-wave thickness for visible light, as mentioned by Richard
Feynman in his famous lectures \cite{Feynman}, and wave plates can
be constructed from this material. Recently, construction of a
quarter-wave retarder for light at $\lambda=632.8$ nm from the 3M
PP2500$^{TM}$ polyester film was proposed \cite{optmat,spie} and
some properties of this film were investigated. While this film is
inexpensive, it is not commonly found in a laboratory. In this paper
we explore the possibility of using most common inexpensive overhead
transparencies of different types (black and white, color, from
different companies) as wave plates and test their quality in
retarder applications. Apart from general interest in a broad class
of applications, this project is also motivated by the need for
cheap wave plates for building commercial and experimental optical
atomic magnetometers based on optical pumping with circularly
polarized light and detection of spins using light polarization
measurements \cite{Allred}. In magnetometer applications, large-size
waveplates are often needed, while the price grows linearly with the
wave-plate area as illustrated in Fig.\ref{price}.

Overhead transparencies are made of polymer film produced by a
process in which the film is extruded through a slit onto a chilled
polished turning roll where it is quenched from one side. Additional
film stretching can be employed for achieving desirable properties.
Such stretching may produce biaxial materials, such as
biaxially-oriented polyethylene terephthalate (boPET) film from
which some overhead transparencies are made for copiers or laser
printers. Refractive indices of boPET were studied in
Ref.~\cite{Elman}. The extrusion and stretching processes in general
lead to biaxial birefringence; however, it is possible that one
preferential direction of molecular orientation is defined and one
refractive index is more pronounced than the others and the medium
can be approximated as uniaxial. For example, in Ref.\cite{Ratta},
the structural properties of HDPE (High Density Polyethylene) film
have been studied and it was found that draw ratio affected
crystalline orientation and refractive indices approximately in the
same way in two directions but quite differently in the third
direction. Birefringence is a very general property of transparent
polymers, including overhead transparencies. The theory of
birefringence induced by mechanical deformation for polymers and
liquid crystals is given in Ref.\cite{Stein}. Optical properties are
of great practical interest because they can be used to analyze the
material structure and to control the extrusion process.
\section{Polarization properties of overhead transparencies }

We investigate polarization properties of overhead transparencies
with a commercial polarimeter, Thorlabs PAX5710IR1-T, which outputs
the state of polarization in the form of  a polarization ellipse
with numerical values of the ellipticity $\chi$ and the azimuth
$\psi$ as well as the Poincar\'{e} sphere with coordinates that are
the three Stokes parameters $s_1$, $s_2$, and $s_3$. The Stokes
parameters can be found from the ellipticity and azimuth, or from
the set of Jones vector parameters such as field amplitudes $a_x$
and $a_y$ and phase shift $\delta$, or vice versa
\cite{bookbornwolf}. For example, the ellipticity and azimuth are
related to the Jones vector parameters through the following
equations
\begin{equation}
\chi=\frac{1}{2}\arcsin\frac{2a_x a_y \sin\delta}{a_x^2+a_y^2},
\end{equation}
\begin{equation}
\psi=\frac{1}{2}\arctan\frac{s_2}{s_1}=\frac{1}{2}\arctan\frac{2 a_x
a_y \cos\delta}{a_x^2-a_y^2}.
\end{equation}
It is convenient to parameterize normalized field amplitudes as
$a_x=\cos\theta$ and $a_y=\sin\theta$, so ellipticity and azimuth
depend only on two variables,
\begin{equation}
\chi=\frac{1}{2}\arcsin[\sin2\theta \sin\delta], \label{chigen}
\end{equation}
\begin{equation}
\psi=\frac{1}{2}\arctan[\tan2\theta \cos\delta].
\end{equation}
Equation \ref{chigen} will be used throughout the paper to relate
the polarimeter's output to a more fundamental property of optical
material such as retardation and to fit experimental data in which
$\theta$ or $\delta$ are varied.

 First, to verify that transparency can operate
as a wave plate, we study the dependence of ellipticity on the
transparency rotation angle. The experimental arrangement for this
measurement is simple: a helium-neon laser beam (632.8 nm
wavelength), linearly polarized with a Polaroid film, is sent to the
detection head of the polarimeter through a fragment of transparency
attached to a rotation mount with angular resolution of 1 degree.
The rotation is defined in Fig.\ref{indell}, left panel. From this
definition if transparency behaves as a waveplate the ellipticity of
the transmitted light as a function of the transparency rotation
must satisfy Eq. (\ref{chigen}) in which $\theta$ is the
transparency rotation angle. The ellipticity of the light as a
function of this angle  is shown in Fig.\ref{transprot}. The
behavior is periodic as expected for a wave plate, in accordance
with Eq. (\ref{chigen}), but some deviation from periodicity is
observed which can be explained by the shift of the
 point where the laser beam intersects the transparency when it is
 not exactly in the center of rotation
combined with spatial variations in the retardation that will be
discussed later. The retardation $\delta$ is about 0.91 rad and
because in the Taylor expansion of arcsine the first term is much
larger than the others, we obtain that the dependence on the
transparency rotation angle is dominantly a sine wave as is seen in
Fig.\ref{transprot}. The maximum value of ellipticity in this
particular measurement is less than maximum possible 45 degrees of a
$\lambda/4$ plate. For constructing a $\lambda/4$ plate a more
appropriate part of the transparency has to be chosen or the plane
of the transparency has to be tilted.

The next graph, Fig.\ref{trans1pos} demonstrates how the position of
the transparency for a fixed orientation affects the induced
ellipticity and hence retardation. Apart from demonstrating how the
retardation can be optimized by spatial selection, this graph also
gives information about typical retardation spatial non-uniformity
in one dimension which is important for the determination of the
useful size of the transparency wave plate. From the data, we see
that the size is of order 1 cm. By using a computer screen
 and a crossed polaroid film it is possible to
find regions with larger area of uniformity as demonstrated in
Fig.\ref{comptr1}.

To understand the order of the transparency waveplate and to
estimate the useful wavelength range where the transparency can
function as a waveplate of a given type, we measured the wavelength
dependence of the ellipticity of a transparency using a tunable
Ti:Sapphire laser as light source (Fig.\ref{transwavelength}). For a
particular angle $\theta=45^\circ$, chosen in the experiment to give
maximum ellipticity, Eq. (\ref{chigen}) predicts the dependence of
ellipticity on retardation as a triangle wave with an amplitude
45$^\circ$. Because the retardation changes approximately linearly
with wavelength in some range
\begin{equation}
\delta=\frac{2\pi\Delta n(\lambda) l}{\lambda}\approx
\frac{2\pi\Delta n(\lambda_0) l}{\lambda_0}- \frac{2\pi\Delta
n(\lambda_0) l (\lambda-\lambda_0)}{\lambda_0^2},
\end{equation}
the ellipticity follows a quasi-periodic triangle wave as a function
of wavelength as observed experimentally. The distance between zeros
can be used to obtain total retardation.

\begin{figure}
\centerline{\includegraphics*[scale=1]{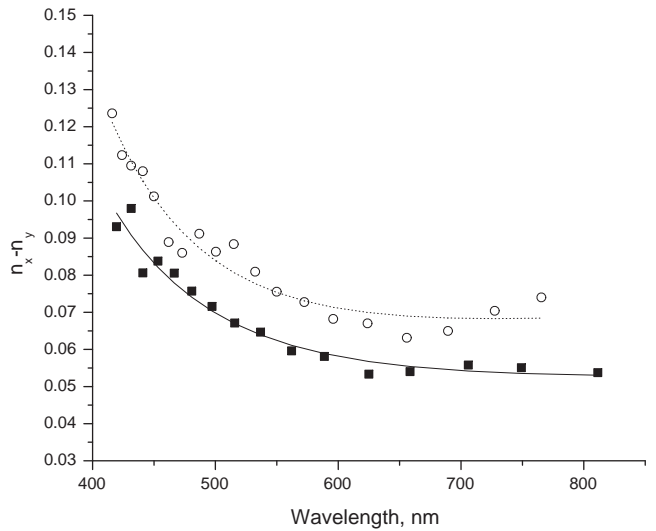}} \caption{
Refractive index difference $n_x-n_y$ of 3M black and white
transparency (circles and dotted line) and color transparency (solid
squares and line) measured with a spectrometer and two crossed
polarizers with a transparency in between. The dotted line is the
fit with $n_x-n_y=0.0806-13100/\lambda^2+3.5\times 10^9/\lambda^4$,
and the solid line is the fit with
$n_x-n_y=0.05628-5500/\lambda^2+2.2\times 10^9/\lambda^4$. }
\label{bwcspectr}

\end{figure}

It is necessary in general to have the dependence of refractive
index on wavelength to obtain total retardation. Such dependence is
often approximated with the Cauchy dispersion formula
$n(\lambda)=n_0+n_1/\lambda^2+n_2/\lambda^4+...$ which is accurate
when wavelength of the light is far from resonances. For wavelengths
shown in the Fig.\ref{transwavelength} the constancy of refractive
index is a good approximation. In a larger range, from 300 nm to 900
nm, this assumption is not satisfied since a strong absorption band
is located near 300 nm. We used a Varian CARY219 spectrometer to
investigate the refractive index variation in the range 400-800 nm.
A transparency film was inserted between two crossed polaroid films
with its x axis at 45$^\circ$, so the transmission depended on the
retardation. The wavelength of minima (or maxima) of the transmitted
intensity could be used to calculate the refractive index difference
$n_x-n_y$ as a function of wavelength. The results are plotted in
Fig.\ref{bwcspectr}. Below 600 nm the wavelength dependence of
refractive indices becomes significant, and a good fit requires
retaining $1/\lambda^2$ and $1/\lambda^4$ terms as it is shown in
Fig.\ref{bwcspectr}.
\begin{figure}
\centerline{\includegraphics*[scale=1]{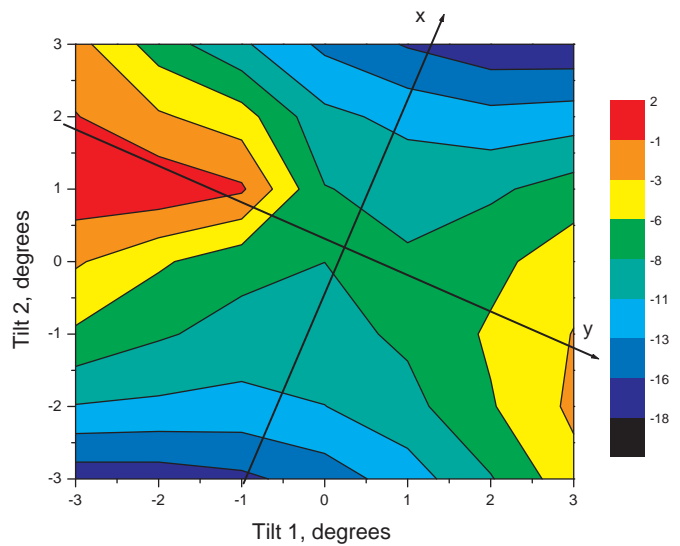}} \caption{The
dependence of ellipticity in degrees on the tilt in two directions
(first around the vertical axis [Tilt 1] and then around the
horizontal axis [Tilt 2]) for 3M black and white transparency. The
ellipticity was maximized by rotating the transparency around its
normal. } \label{twodimtilt}
\end{figure}
\begin{figure}
\centerline{\includegraphics*[scale=1]{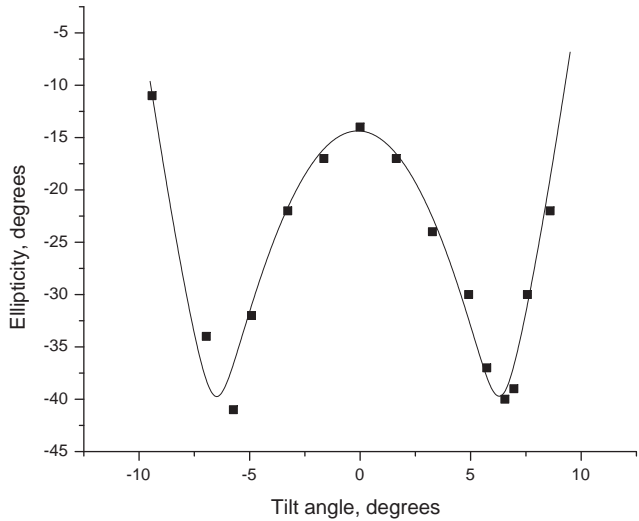}}
\caption{Dependence of ellipticity on the tilt angle $t$ of a color
transparency sheet for the tilt around a principal axis of the
refractive index ellipsoid. Solid line is the fit based on Eqs.
(\ref{chigen}) and (\ref{downpar}).} \label{tiltfast}
\end{figure}

It is interesting to note that a transparency between two polarizers
can serve as a birefringent filter. Investigation of such filters
based on low-cost polymers has been carried out in
Ref.\cite{Velasquez} where filters with bandwidth 50 nm were
demonstrated.

It might be convenient to adjust the retardation to a desirable
level by tilting a transparency waveplate. The dependence on the
tilt is determined by two factors: one is the increase in the
optical path, and the other is the change in the retardation per
unit length. In the general case of a biaxial medium the dependence
of retardation and ellipticity on the tilt angle is quite
complicated and for the analysis it is important to find the
principal axes of the index ellipsoid, see Fig.\ref{indell}. To
first approximation one in-plane principal axis, x or y, with the
assumption to be justified in a moment that z is normal to the
transparency, can be found easily by determining tilt direction for
which the change in retardation for a small tilt angle
($\pm3^\circ$) is the largest. From the dependence of retardation on
the tilt angle (Eq.\ref{downpar}) for this tilt direction we
estimated that $n_z-n_x$ was about 0.2, much larger than
$n_x-n_y\approx0.06$ obtained from the wavelength measurements at
normal incidence (Fig.\ref{bwcspectr}) for the transparency of this
type. Such small retardation for normal incidence means that the z
axis of the index ellipsoid has a relatively small projection in the
transparency plane and hence our zero-approximation assignment of
refractive index subscripts is justified. After finding the z axis
approximately, we can determine the principal refractive indices
more accurately by investigating the retardation for small
deviations of the transparency film normal from the z direction. If
$n_a$ and $n_b$ are defined as the semi-axes of the ellipse obtained
by the intersection of the plane normal to the $\bf{k}$ vector with
the index ellipsoid (Refs.\cite{bookbornwolf,bookPhotonics}), the
retardation is proportional to $n_a-n_b$. For tilt of the ellipsoid
around the x axis (Ref.\cite{bookbornwolf}, section 14.3.3),
\begin{equation}
n_a-n_b=n_x-n_y-(n_z-n_y) \sin(t_x/n)^2 \label{downpar}
\end{equation}
($t$ is the tilt angle, $n$ is the average refractive index), and
the retardation decreases with tilt if $n_x<n_y<n_z$ while for the
tilt around the y axis,
\begin{equation}
n_a-n_b=n_x-n_y+(n_z-n_x)\sin(t_y/n)^2
\end{equation}
the retardation increases with tilt.  For small tilt, in the first
case the dependence is a downward parabola symmetric with respect to
zero tilt; in the second case the dependence is an upward parabola.
Generally, the dependence of retardation on the tilt has a saddle
point with orthogonal symmetry axes. If two dimensional data are
given, the direction of the principal axis z can be found as normal
to the surface going through the saddle point, and that of x and y
axes can be found as the lines of symmetry. In Fig.\ref{twodimtilt},
for example, the x axis constitutes about $21^\circ$ angle with
horizontal axis. If this angle is known, then from the fitting for
tilt is sufficient to obtain $n_z-n_y$ or $n_z-n_x$. If we neglect
small effect due to path elongation, then $n_z-n_x\simeq 0.23$,
close to our initial estimate.

Different transparencies have slightly different properties,
although qualitatively they are similar, including the orientation
of the index ellipsoid. In Fig.\ref{twodimtilt} we show tilt
measurements for a 3M black and white transparency. To give an idea
of variations, in Fig.\ref{tiltfast} we also present tilt
measurements for the color transparency, with direction of the tilt
around the x axis.  This graph also illustrates the behavior of
ellipticity in a larger tilt range and shows that 45$^\circ$
ellipticity can be achieved by tilting at a relatively small angle
of 7$^\circ$. The dependence in Fig.\ref{tiltfast} is quite
symmetric in accordance to the previously drawn conclusion that the
z axis is orthogonal to the plane as depicted in Fig.\ref{indell} on
the left panel. We found using crossed polarizers that the x axis
makes a constant angle about $20^\circ$ with respect to the edges of
the transparency sheet across a large transparency area, although it
varies for different samples. Using parabolic fits, we obtain
$n_z-n_y\simeq0.18$, again much larger than $n_x-n_y=0.06$ in
accordance with the z axis perpendicular alignment to the
transparency plane. We have qualitatively investigated the
dependence of ellipticity of this transparency on the tilt angle for
different tilt direction, around x, y, and the xy bisector. The
results are consistent with theoretical predictions: for the tilt
around x - the dependence is a downward parabola, around y - upward,
and for the tilt around xy the dependence is very slow (see
Fig.\ref{twodimtilt}).

To complete the analysis it was also necessary to know the average
refractive indices. We have measured them using Snell's law by
sending a 633-nm light beam at the edges of a stack of
transparencies and by tracing incident and refracted rays. Several
measurements for different incidence angles resulted in the average
value $n=1.54\pm0.05$. For comparison similar values at the same
wavelength are obtained in Ref.\cite{Elman} for biaxially-stretched
poly-ethyleneterephthalate (PET) films, $n=1.59$, and in
Ref.\cite{Martinez} for polyester biaxial stretched films, $n=1.66$,
while measurements in Ref.\cite{spie} for 3M PP2500 films gave
different result, $n\approx1.83$. Such a high refractive index is
uncommon for transparent organic materials which consist mostly of
carbon and hydrogen, with single molecular bonds, and have similar
densities. In order to check this result we conducted our own
measurements using beam trace method as above for 3M PP2500 film. We
found that for the polarization perpendicular to the plane of the
sheet, $n=1.43-1.58$ (the spread is due to both inaccuracy and
birefringence) and for the polarization in the plane of the sheet
$n=1.34-1.65$ with error of measurements about 0.05, in disagreement
with Ref.\cite{spie}.

It is also interesting to compare the results for the differences
$n_i-n_j$. Our results for two different transparency types are
$n_z-n_x\approx 0.2$ and $n_x-n_y\approx 0.05$. In Ref.\cite{Elman}
for similar definition of principal axes, $n_x-n_z\approx0.15$ and
$n_y-n_x\approx0.011$, roughly in agreement. By interpreting results
of Ref.\cite{spie}, the differences are of similar order. In
general, birefringence depends strongly on the details of
manufacturing process, so only qualitative agreement is expected.

\section{Applications}
The simplest method to construct a desired wave plate for a given
wavelength is to use a light source of this wavelength and to choose
an appropriate area in a transparency sheet, monitoring retardation
with either a commercial polarimeter or some other optical
polarimetric arrangement. On the other hand, tuning can be achieved
by tilting. Wavelengths for which transparency retarders can be
constructed are in the wide range from visible to infrared. In
particular, we tested operation at 633 nm with a helium-neon laser,
at 790-930 nm with a Ti:Sapphire laser, and in the range of 300-900
nm with a spectrometer, as well as using a computer monitor with
pure color settings.

\begin{figure}
\centerline{\includegraphics*[scale=0.9]{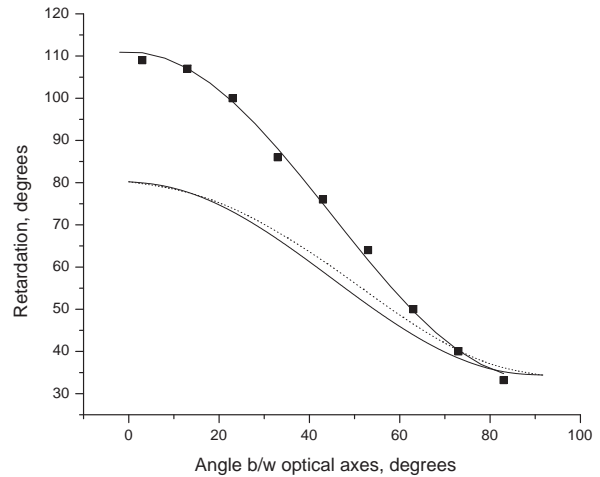}}
\caption{Dependence of retardation of a combination of two wave
plates on the angle between their symmetry axes. Points are
experimental data for black\&white 3M transparencies. They allow
accurate fit with analytical equation Eq. (\ref{inteq}), a solid
line, with $\phi_2=72.3$ degrees and $\phi_1=38.7$ degrees.
Measurements for individual wave plates gave $\phi_2=74.6$ degrees
and $\phi_1=37$ degrees, in agreement. Small deviation is due to
spatial variations of retardation when transparency is rotated.
Analytical equation for retardations $\phi_1=23$ degrees and
$\phi_2=57$ degrees, another solid line, is also tested by
comparison with numerical simulations based on the Jones-matrix
formalism, dotted line.} \label{waveplatecombo}
\end{figure}

 In addition, we have also investigated the possibility of
superposition of two transparency sheets for achieving desirable
retardation. Since the eigenvectors of the Jones matrix of a
combination of two waveplates are in general complex, the system has
elliptical-polarization eigenstates and is not equivalent to a
single waveplate which has eigenstates of linear polarization,
except in some degenerate cases. In other words, the
two-transparency combination can be viewed as a waveplate combined
with a gyrotropic plate, whose effect is just a rotation of the
polarization ellipse. If the purpose of the waveplate is, for
example, to produce a polarization state with a given ellipticity
starting  with linearly polarized light, then a combination
waveplate can generally be used for this. From the Jones-matrix
formalism one finds that a two-transparency combination produces the
same output ellipticity as a single waveplate, which has retardation
$\phi$ between $\phi_1-\phi_2$ and $\phi_1+\phi_2$ depending on the
angle between the fast axes of the two wave plates $\theta_{12}$.
For practical purposes this dependence can be interpolated with
sufficient accuracy with
\begin{equation}
\phi=(\phi_1+\phi_2)\cos^2\theta_{12}+(\phi_1-\phi_2)\sin^2\theta_{12}
\label{inteq}.
\end{equation}
A comparison of numerical simulations with this interpolation
equation for a typical case is shown in Fig.\ref{waveplatecombo}. To
confirm this theory, we investigated experimentally the behavior of
two-sheet waveplates by rotating them and by varying the angle
between the axes of the two constituent transparencies. First we
tested that the ellipticity of two transparencies under the common
rotation follows Eq. (\ref{chigen}). Second, from the fits by Eq.
(\ref{chigen}) we found retardation for different relative
orientations of the transparencies. Figure \ref{waveplatecombo}
summarizes the analysis and shows that the interpolation equation
(Eq. (\ref{inteq})) agrees with the experimental data.
In principle, even a single transparency sheet may have gyrotropy.
Experimentally, we found that a single sheet of transparency does
not change polarization if input polarization is aligned with the
optical axis; so a single transparency behaves as a true
birefringent wave plate.

It is interesting to note that when the angle between the fast axes
of two transparency sheets is 90$^\circ$, the combined retardation
will be lower, meaning that the order is lower, and the sensitivity
to variations in parameters, such as temperature and wavelength,
will be also lowered due to cancelation. This is a common technique
for building low-order wave plates and is described for example in
Ref.\cite{bookopt3}.

Although the fast wavelength  dependence of the retardation for a
single sheet would limit the range of operation of a transparency
wave plate to $\sim$10 nm (Fig.\ref{transwavelength}), a combined
plate can work in a much larger range.

\begin{figure}
\centerline{\includegraphics*[scale=1]{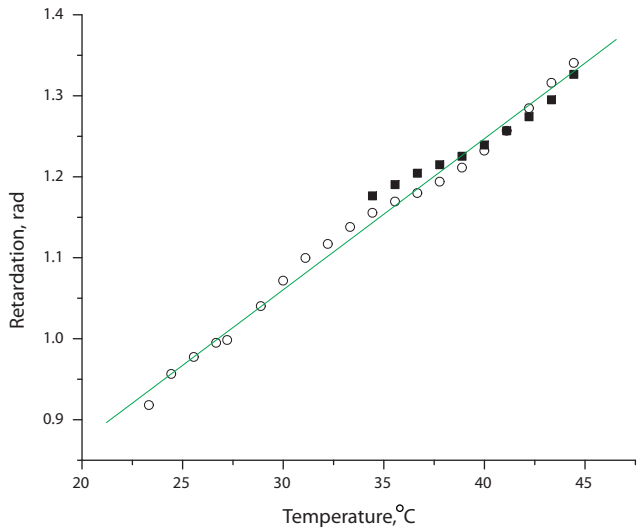}}
\caption{Dependence of retardation of a color transparency on
temperature at 633 nm. Open circles show a set of data in one run,
and the squares the second run after heating-cooling cycle to check
reproducibility of the data. The dependence fits well with a line,
$\phi=0.50(1)+0.0187(4)T$, where $T$ is the temperature. Large
offset $\sim20\pi$ rad which can be estimated from refractive index
for the color transparency (Fig.\ref{bwcspectr}) is not included. }
\label{temprdepend}
\end{figure}

One additional possibility is to combine a high-quality wave plate,
which has some retardation offset due to operation at a non-nominal
wavelength, with a two-transparency wave plate to compensate the
offset. If we choose a 90-degree relative orientation in the
combined transparency wave plate, the temperature stability of the
resulting wave plate will not be significantly reduced.

Similarly, although the temperature dependence of  the retardation
of a transparency is substantial, as shown in Fig.\ref{temprdepend},
the combined wave plate has a much weaker temperature dependence.
The same is also true for the more conventional composite low-order
optical, for example, crystalline quartz, plates \cite{bookopt3}.
This dependence is linear in a wide range, which can be explained by
linear thermal expansion. Assuming that the expansion is isotropic
and $\Delta l/l=\alpha \Delta T$, where $\alpha$ is the coefficient
of linear expansion and $\Delta l/l$ is the relative change in the
length for temperature increment $\Delta T$, the density is reduced
by $1/(1+\alpha \Delta T)^3$ and the total retardation is reduced by
$1/(1+\alpha \Delta T)^2\approx 1-2 \alpha \Delta T$, where we have
taken into account the change in the plate's length. From the total
retardation of the transparency of the order $20\pi$ rad (for
$n_x-n_y=0.05-0.06$, [Fig.\ref{bwcspectr}]), we can estimate the
coefficient of linear expansion to be $\alpha\approx1.5\times
10^{-4}$ $K^{-1}$, in agreement with the common values for such
materials, Ref.\cite{physchemconstref}. Apart from being a parasitic
effect, the temperature dependence can be used in some applications.
For example, it is possible to construct multi-channel thermometers
based on monitoring retardation using broad light beams. Another
application is the conversion of UV or IR images into visible ones
by measuring retardation across the transparency (the local
retardation is affected by heating with the absorbed UV or IR
light). For small sizes, the transparency retardation is uniform and
imaging can be implemented directly, but for large sizes
non-uniformity can be taken into account with a computer. It is
interesting to note that using sensitive polarimeters with
fundamental photon shot noise on the order of 10$^{-9}$
rad/Hz$^{1/2}$, it should be possible to measure temperature changes
as small as $10^{-7}$ Kelvin with a single sheet and even better
using a stack of transparencies. Other parameters such as ambient
pressure or electric field (Kerr effect), can be measured with high
sensitivity. Electric field imaging can be also of interest.

The wavelength dependence of the retardation
(Fig.\ref{transwavelength}) can be also used for sensitive
wavelength measurements. For example, using again shot-noise limit
of order 10$^{-9}$ rad/Hz$^{1/2}$ in angular resolution, we estimate
that changes in wavelength on the order of $3\times 10^{-8}$ nm can
be detected for $\sim$1 second of integration time.

 In conclusion, we have analyzed polarization
properties of overhead transparencies and found that they can be
used as inexpensive wave plates in a broad range of wavelengths.

\section{Acknowledgment}
This work is supported by DOD MURI grant \# N-00014-05-1-0406. The
authors are grateful to A. Cingoz and D. English for helping with
measurements of wavelength dependence, and to A. Sushkov, A. Park,
J. Higbie, T. Karaulanov, and M. Ledbetter for useful comments on
the manuscript.
\bibliography{transp}
\end{document}